\documentclass[runningheads]{llncs}

\usepackage{multirow}
\usepackage{marvosym}

\usepackage{graphicx}
\usepackage[caption=false]{subfig}

\usepackage{makecell}
\usepackage{booktabs}

\usepackage{array}
\newcommand{\PreserveBackslash}[1]{\let\temp=\\#1\let\\=\temp}
\newcolumntype{C}[1]{>{\PreserveBackslash\centering}p{#1}}

\begin{document}              % start of a contribution

\title{Uptrendz: API-Centric Real-time Recommendations in Multi-Domain Settings}
\titlerunning{API-Centric Real-time Recommendations in Multi-Domain Settings} 
%

%%
%% The "author" command and its associated commands are used to define
%% the authors and their affiliations.
%% Of note is the shared affiliation of the first two authors, and the
%% "authornote" and "authornotemark" commands
%% used to denote shared contribution to the research.

\author{Emanuel Lacic \inst{1} \and Tomislav Duricic \inst{1,2} \and Leon Fadljevic \inst{1} \and \\ Dieter Theiler \inst{1} \and Dominik Kowald$^{\textrm{(\Letter)}}$ \inst{1,2}}
\authorrunning{E. Lacic, T. Duricic, L. Fadljevic, D. Theiler, and D. Kowald}

\institute{Know-Center GmbH, Graz, Austria \\\email{\{elacic,tduricic,lfadljevic,dtheiler,dkowald\}@know-center.at} \and
Graz University of Technology, Graz, Austria}

\maketitle

\begin{abstract}
In this work, we tackle the problem of adapting a real-time recommender system to multiple application domains, and their underlying data models and customization requirements. To do that, we present Uptrendz, a multi-domain recommendation platform that can be customized to provide real-time recommendations in an API-centric way. We demonstrate (i) how to set up a real-time movie recommender using the popular MovieLens-100k dataset, and (ii) how to simultaneously support multiple application domains based on the use-case of recommendations in entrepreneurial start-up founding. For that, we differentiate between domains on the item- and system-level. We believe that our demonstration shows a convenient way to adapt, deploy and evaluate a recommender system in an API-centric way. The source-code and documentation that demonstrates how to utilize the configured Uptrendz API is available on GitHub.


\end{abstract}

\section{Introduction}
Utilizing recommender systems is nowadays recognized as a necessary feature to help users discover relevant content \cite{resnick1997recommender,mcnee2006being}. Most industry practitioners~\cite{amatriain2016past}, when they build a recommender system, adapt existing algorithms to the underlying data and customization requirements of the respective application domain (e.g., movies, music, news, etc.). However, the focus of the research community has recently shifted towards building recommendation systems that simultaneously support multiple application domains \cite{bonab2021cross,im2007does,roitero2020leveraging} in an API-centric way.

In this work, we demonstrate Uptrendz\footnote{\url{https://uptrendz.ai/}}, an API-centric recommendation platform, which can be configured to simultaneously provide real-time recommendations in an API-centric way to multiple domains. Uptrendz supports popular recommendation algorithms, e.g., Collaborative Filtering (CF), Content-based Filtering (CBF, or Most Popular (MP), that are applied across different application domains. The focus of this demonstration is to show how domain-specific data-upload APIs can be created to support the customization of the respective recommendation algorithms. Using the MovieLens-100k dataset \cite{harper2015movielens} and a real-world use-case of entrepreneurial start-up founding\footnote{\url{https://cogsteps.com/}}, we show how such an approach allows for a highly customized recommendation system that can be used in an API-centric way. The source-code and documentation for this demonstration is available via GitHub\footnote{\url{https://github.com/lacic/ECIR2023Demo}}.

\begin{figure}[t!]
  \centering
\includegraphics[width=0.99\textwidth]{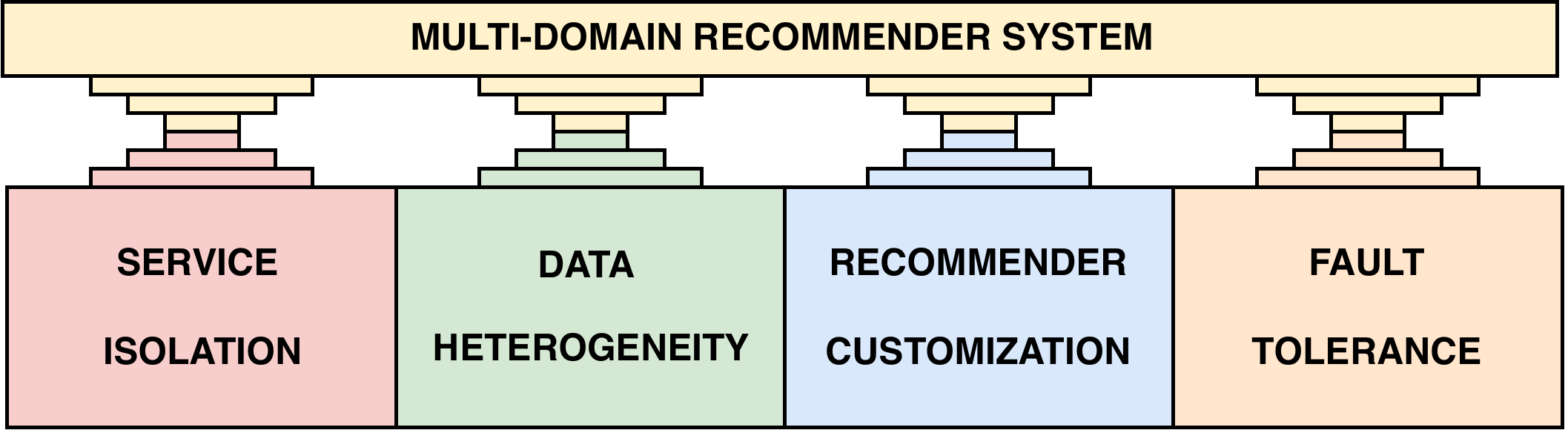}
    %\vspace{2mm}
      \caption[Multi-domain design aspects]{Aspects that need to be addressed when building a recommender system for a multi-domain environment \cite{lacic2017tailoring}.}
   \label{fig:multi_domain_guidelines}
\end{figure}

\section{The Uptrendz Platform}
The Uptrendz platform is built on top of the ScaR recommendation framework \cite{lacic2014towards}. As shown in \cite{lacic2017tailoring} and Figure \ref{fig:multi_domain_guidelines}, the microservice-based system architecture addresses four distinctive requirements of a multi-domain recommender system, i.e., (i) service isolation, (ii) data heterogeneity, (iii) recommender customization, and (iv)
fault tolerance. 
Uptrendz provides a layer on top of the framework to dynamically configure an application domain and to instantly provide an API to (i) upload item, user and interaction data, and (ii) request recommendations. 

\vspace{2mm}
\noindent
\textbf{Domain-specific data model.} As discussed by \cite{Adomavicius2010}, different domains may employ the same recommender algorithm but can differ with respect to what kind of data is utilized to build the model (e.g., interaction types, context, etc.).  
Given an API-centric approach, we show that in order to support the customization of recommender algorithms with domain-specific parameters, the underlying platform needs to unambiguously know which source of information should be used to calculate the recommendations. To do that, the Uptrendz platform first allows generating customized data upload APIs for multiple item and user entities (see Table \ref{tab:data}). Second, with respect to interaction data, both user-item and user-user interactions can be configured. The interaction API is further customized in accordance to what kind of interactions the respective application domain actually supports, i.e., (i) registered users, anonymous sessions or both, (ii) interaction timestamp tracking, and (iii) type of interaction (explicit or implicit).

\vspace{2mm}
\noindent
\textbf{Recommender customization.} The Uptrendz platform fosters the notion of defining personalization scenarios (i.e., use-cases) when creating recommendation APIs. The available selection of real-time recommendation models \cite{lacic2014towards} for a given scenario depends on (i) what should be recommended (e.g., item or user entities), (ii) for whom the recommendations are targeted (e.g., registered or anonymous users) and, (iii) what kind of context is given \cite{adomavicius2011context} (e.g., item ID to recommend relevant content for). As we adopt a non-restricted configuration with respect to the number of freely defined user interaction types, algorithms that use this kind of data (e.g., Collaborative Filtering) can be customized to utilize any subset of the list of available interactions as well as to define how much weight a particular interaction type should have. With respect to post-filtering recommendation results, each model can use categorical (e.g., tags~\cite{lacic2014recommending} or other semantic representations~\cite{kowald2013social}) or numerical data attributes to ensure that the resulting recommendations either contain or exclude a particular value (see Table \ref{tab:data} for complete list of attributes). 

\begin{table}[!t]
    \centering
        \caption{Supported attributes to configure the data upload API for items and users.}
    \begin{tabular}{C{0.15\linewidth}|C{0.14\linewidth}||p{0.65\linewidth}}
        \thead{ Type} & Sub-Type & Description \\ \hline\hline
         \multirow{4}{*}{\parbox{\linewidth}{\centering Categorical Text}} &  \multirow{2}{*}{\parbox{\linewidth}{\centering Single Value}} & 
         
         String value, which usually represents a category. Used for \textbf{post-filtering} recommendation results. \\ \cline{2-3}
         & \multirow{2}{*}{\parbox{\linewidth}{\centering Multiple Values}} & List of string values, which usually represent an array of categories. Used for \textbf{post-filtering} recommendation results. \\ \hline
        \multirow{4}{*}{Free Text} & \multirow{2}{*}{English} & \textbf{English text}, which is processed and utilized for \textbf{content-based} recommendations. \\ \cline{2-3}
         & \multirow{2}{*}{German} & \textbf{German text}, which is processed and utilized for \textbf{content-based} recommendations. \\ \hline
        \multirow{2}{*}{Numeric} & \multirow{1}{*}{Integer} & Used for \textbf{post-filtering} recommendations (e.g., user age). \\ \cline{2-3}
         & \multirow{1}{*}{Real} & Used for \textbf{post-filtering} recommendations (e.g., price). \\ \hline
        \multirow{1}{*}{Date} & \multirow{1}{*}{-} & Date information for the respective entity (e.g., creation date) \\ \hline
    \end{tabular}
    \label{tab:data}
\end{table}

%%%%%%%%%%%%%%%%%%%%%%%%%%%%%%%%%%%%%%%%%%%
\section{Multi-Domain Support}
In order to provide a multi-domain recommender platform, we support the notions of a system-level and item-level domain in accordance with \cite{cantador2015cross}. For the former, items and users belong to distinct systems (e.g., Netflix and Amazon). For the latter, individual domains have different types of items and users which may share some common attributes (e.g., movies and books).

\vspace{2mm}
\noindent
\textbf{Demo Walkthrough: System-level domain.}
When a domain is created on a system level, the underlying data is physically stored in a different location than the data of other domains. Hence, domains do not share any data between themselves and the underlying services are isolated so that the performance of one domain does not impact the performance of another domain (e.g., during request load peaks). 
We demonstrate how to create a movie recommender on a system level. To utilize the MovieLens-100k dataset~\cite{harper2015movielens}, we first need to configure the respective data services to upload (i) movie, (ii) user, and (iii) interaction data. Each entity needs to be separately created in the Uptrendz platform in order to generate an API that can be used to upload the MovieLens-specific data attributes. This allows creating recommendation scenarios for (i) similar movies (CBF), (ii) popular horror movies (MP with post-filtering), (iii) movies based on ratings (CF), (iv) their weighted hybrid combination (e.g., for cold-start settings~\cite{lacic2015tackling}, and (v) a user recommender for a given movie.

\begin{figure}[t!]
  \centering
    \includegraphics[width=0.99\textwidth]{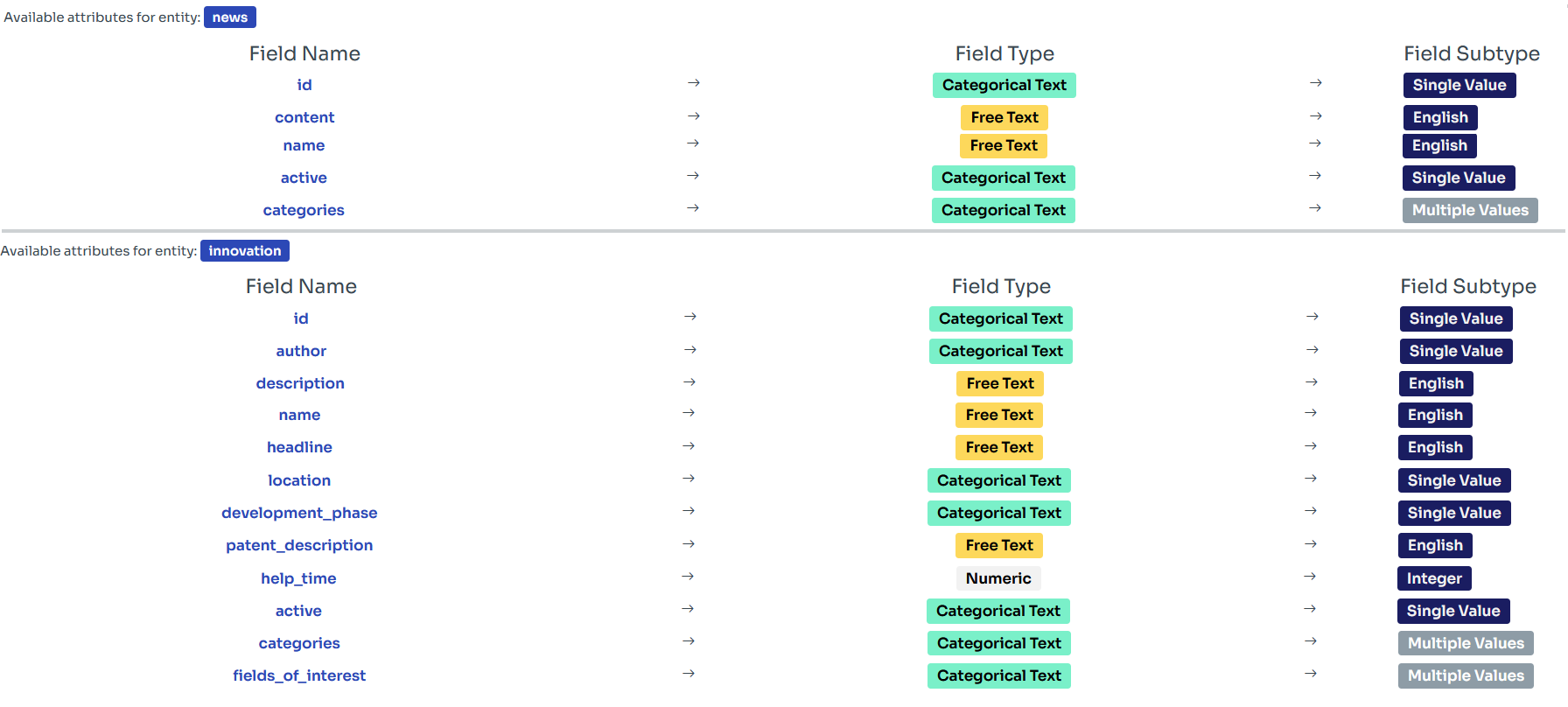}
    \includegraphics[width=0.99\textwidth]{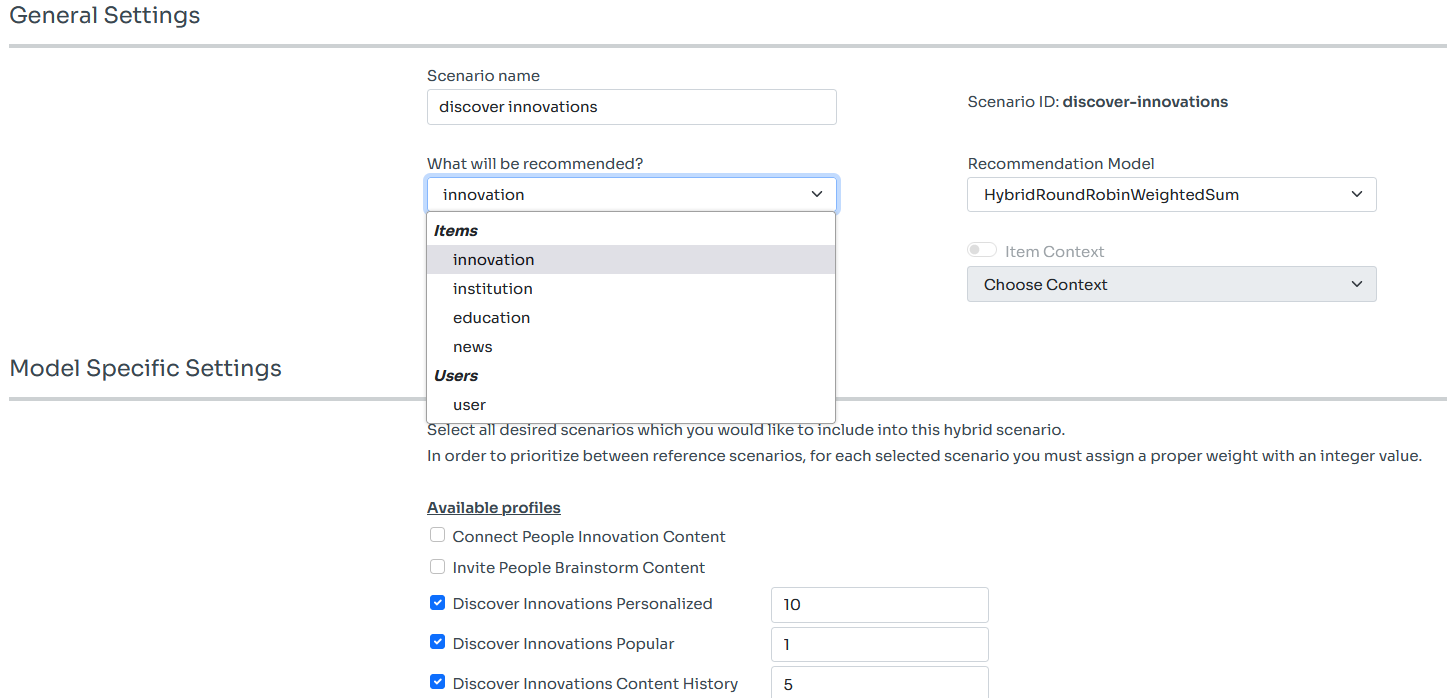}
    %\vspace{2mm}
      \caption{Example of supporting multiple domains on the item-level (up) and configuring a hybrid recommendation algorithm (below) with previously created APIs.}%\vspace{-3mm}}
    \label{fig:cogsteps}
    %\vspace{-2mm}
\end{figure}

\begin{figure}[t!]
  \centering
    \includegraphics[width=0.49\textwidth]{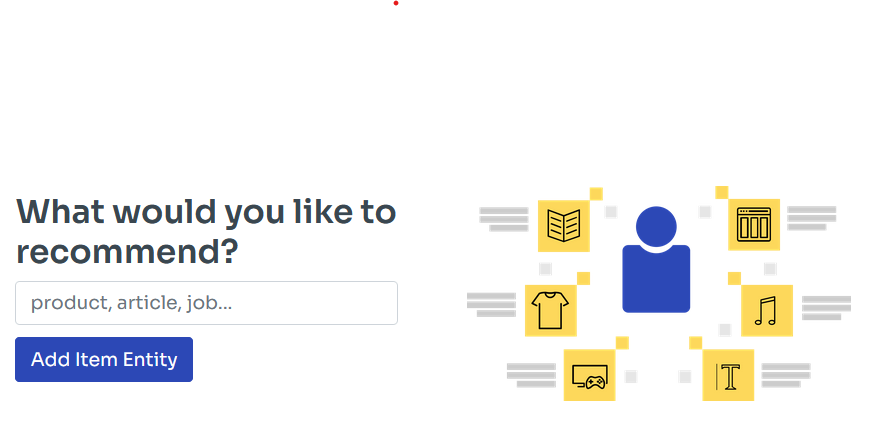}
    \includegraphics[width=0.49\textwidth]{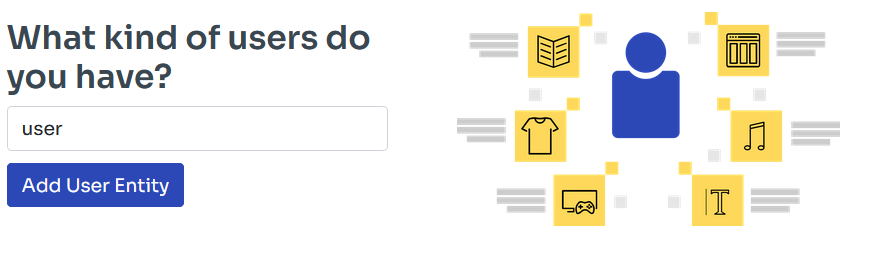}
    %\vspace{2mm}
      \caption{Uptrendz requires the specification of (i) the item types that should be recommended (e.g., products or users, depending on the domain - left figure), and (ii) the user types for which recommendations should be generated (e.g., registered users or session users - right figure).}
    \label{fig:data_entities}
\end{figure}

\vspace{2mm}
\noindent
\textbf{Demo Walkthrough: Item-level domain.}
To showcase how to configure Uptrendz to support multiple-domains on an item-level, we present the use-case of entrepreneurial start-up founding.
Here, we recommend experts that can provide feedback to an innovation idea, support co-founder matching, help incubators, innovation hubs and accelerators to 
discover innovations but also provide relevant educational materials until the innovation idea matures enough to form 
a start-up. In this case, each recommendable entity has a separate data model and can be viewed as part of a standalone application domain. Figure \ref{fig:cogsteps} depicts how adding multiple item entities in the data catalog allows customizing data attributes for the respective domain. While configuring a recommendation algorithm, the respective item-level domain can be selected to be recommended. Here, via the example of a hybrid algorithm, only pre-configured algorithms can be utilized that belong to the same domain (i.e., innovation recommendations). 

Finally, in Figure~\ref{fig:data_entities}, we show how Uptrendz allows the specification of (i) different item types that can be recommended, and (ii) different user types for which recommendations should be generated. Our demo application includes different specification examples.

\section{Conclusion}
In this paper, we present Uptrendz, an API-centric recommendation platform that can be customized to provide real-time recommendations for multiple domains. To do that, we support the notions of a system-level and item-level domain. We demonstrate Uptrendz using the popular MovieLens-100k dataset and the use-case of entrepreneurial start-up founding.

In future work, we plan to support even more use cases from other domains, e.g., music recommendations~\cite{kowald2021support}. Here, we also want to integrate fairness-aware recommendation algorithms for mitigating e.g., popularity bias effects.

\vspace{2mm}
\noindent
\textbf{Acknowledgements.} This research was funded by CogSteps and the ``DDAI'' COMET Module within the COMET – Competence Centers for Excellent Technologies Programme, funded by the Austrian Federal Ministry for Transport, Innovation and Technology (bmvit), the Austrian Federal Ministry for Digital and Economic Affairs (bmdw), the Austrian Research Promotion Agency (FFG), the province of Styria (SFG) and partners from industry and academia. 

%%
%% The next two lines define the bibliography style to be used, and
%% the bibliography file.
\bibliographystyle{splncs04}
\bibliography{refs}

\end{document}